\documentclass[aps,prl,twocolumn,superscriptaddress]{revtex4}

\usepackage{amsmath,amsfonts,amssymb,slashed}
\usepackage[english]{babel}
\usepackage{graphicx}
\usepackage{color}
\usepackage{psfrag}

\begin{document}

\title{Probing the BCS-BEC crossover with photons in a nonlinear optical fiber}

\author{Mingxia Huo}
\affiliation{Centre for Quantum Technologies, National University of Singapore, 2 Science Drive 3, Singapore 117542.}

\author{Changsuk Noh}
\affiliation{Centre for Quantum Technologies, National University of Singapore, 2 Science Drive 3, Singapore 117542.}

\author{B. M. Rodr\'{\i}guez-Lara}
\affiliation{Centre for Quantum Technologies, National University of Singapore, 2 Science Drive 3, Singapore 117542.}

\author{Dimitris G. Angelakis}
\email{dimitris.angelakis@gmail.org}
\affiliation{Centre for Quantum Technologies, National University of Singapore, 2 Science Drive 3, Singapore 117542.}
\affiliation{Science Department, Technical University of Crete, Chania, Crete, Greece, 73100}

\begin{abstract}

We propose a scheme where strongly correlated photons generated inside a hollow-core one-dimensional fiber filled with two atomic species can be used to simulate the BCS-BEC crossover. We first show how stationary light-matter excitations (polaritons) in the system can realize an optically tunable two component Bose-Hubbard model, and then analyze the optical parameters regime necessary to generate an effective Fermi-Hubbard model of photons exhibiting Cooper pairing. The characteristic correlated phases of the system can be efficiently observed due to the {\it in situ} accessibility of the photon correlations with standard optical technology.

%The probing of the characteristic correlated phases of the system is done through releasing the trapped photonic states and measuring the relevant correlation functions using standard quantum-optical technology.

\end{abstract}


\maketitle

%Introduction

Superconductivity is undoubtedly one of the most fascinating and elusive condensed matter phenomena \cite{HighTc}.  Bardeen-Cooper-Schrieffer (BCS) theory \cite{BCS} provided the first satisfying explanation of the effect, by proposing that fermions form long range pairs (Cooper pairs) under an arbitrarily weak attractive interaction. A minimal model exhibiting Cooper pairing is the attractive Fermi-Hubbard (FH) model \cite{Bloch_review,BCS_theory_rest}. 
In the BCS-like region characterized by weak inter-species attraction, large cooper-pair-like states form the ground state and exhibit long-range correlations. The latter is destroyed as the inter-species attraction is raised and localized bosonic molecules are formed.
Seminal experimental realizations of FH model require dilute fermi gases for temperatures well below the degeneration temperature, making these experiments extremely challenging \cite{EXP_FERMIONS}.
%So far the BEC-BCS crossover has been observed in quantum degenerate gases involving two or three fermionic species where the use of strong pairing mechanisms yields superfluid transition temperatures close to achievable experimental regimes; \textit{e.g.} involving two \cite{Holland} or three \cite{Martikainen} fermionic species.

A different approach involves utilizing the well known mapping of one-dimensional hard-core bosons into free spinless fermions \cite{Girardeau}. This led to works suggesting bosonic mixtures on an optical lattice in the regime of strong intra-species repulsion for an effective realization of the necessary interacting fermionic behaviour \cite{Paredes}.
In this case, the so-called BCS-BEC crossover could be observed for a higher temperature and furthermore a new phase appears as the inter-species attraction is increased in comparison to the intra-species repulsion. In the latter regime, the system moves away from the fermionic BCS-BEC regime and enters a new strongly localized bosonic phase termed big boson (BB) with almost all the bosonic molecules occupying the same site.

In this work we show that one could circumvent the issues around bosonic or fermionic atoms and actually use  photons in a nonlinear optical medium to efficiently simulate the crossover. We show how to generate a highly tunable two component Bose-Hubbard (BH) model of polaritons or photons  in a nonlinear fiber and analyze the probing of BCS-BEC crossover using optical methods. Our work is motivated by recent progress in the field of photonic quantum simulations \cite{SIPS}, where photon crystallization,  Luttinger liquids behaviour and the ``Pinning transition" has been recently predicted in hollow core optical fibers \cite{hollow_recent}. 

In the following, we introduce a possible experimental setup and analyze the conditions for the realization of a two component BH model of photons.  We then investigate the possibility of tuning the photonic intra-species repulsion to the necessary regime to generate effective fermionic behaviour with attractive inter-species interactions. Our photonic system is then shown to exhibit the BB-BEC-BCS crossover as the optical parameters are varied within realistic regimes. Finally, we discuss a possible observation of the crossover by analyzing specific correlation functions of the released photon pulses as they exit the fiber.
%%%%%%%%%%%%%%%%%%%%%%%%%%%%%%%%%%%%%%%%%%%%%%%%%%%%%%%%%%%%%%%%%%%%%%%%%%%%%%%
\begin{figure}
\includegraphics[width=2.5in]{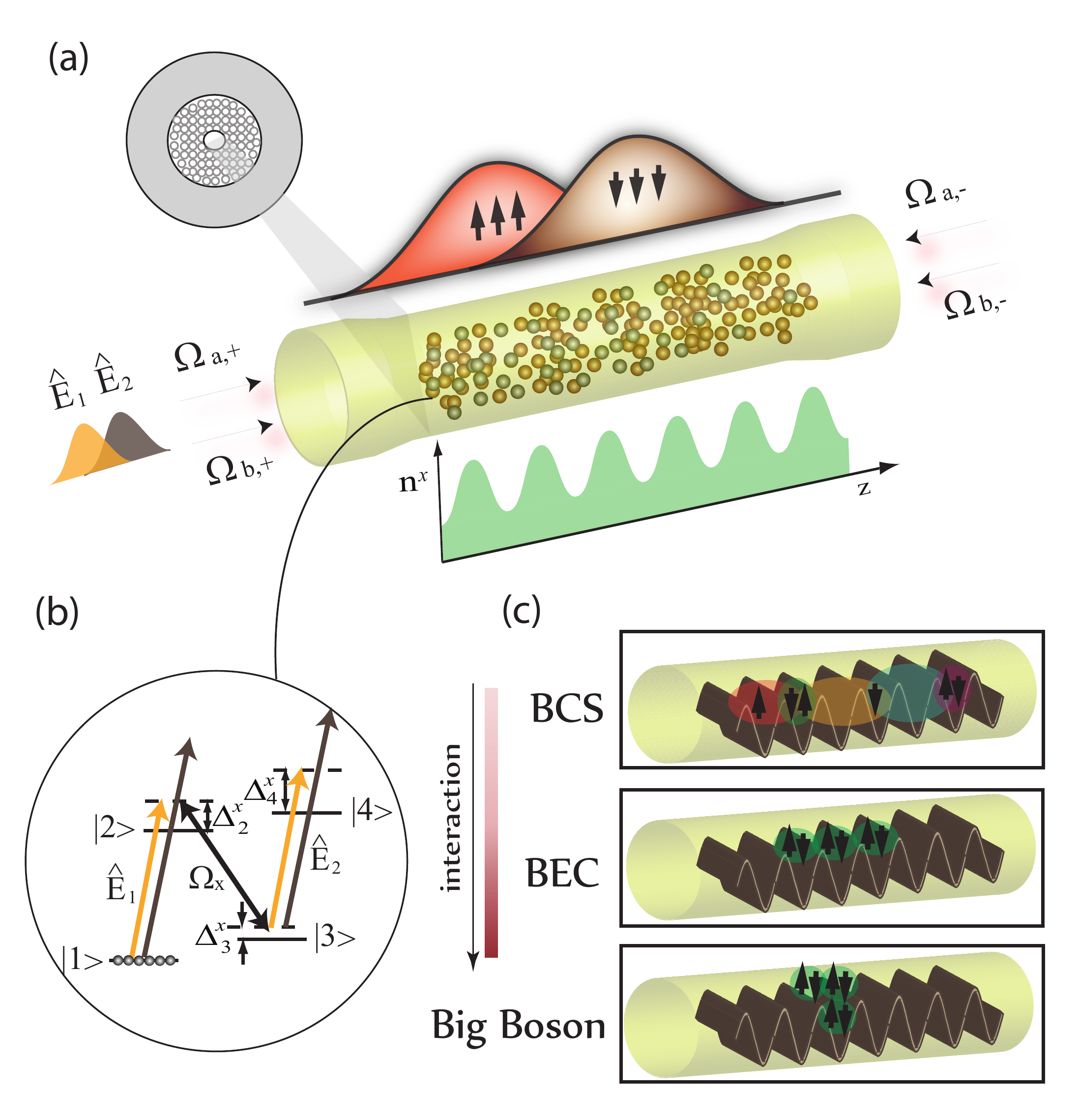}
\caption {(a) A schematic diagram of the system under study. A hollow-core optical waveguide is filled with a mixture of two species of cold atoms interacting with a pair of quantum, $\hat{E}_{(1,2)}$, and a pair of classical, $\Omega _{(a,b)}$, light fields. The resulting stationary light-matter excitations in the waveguide can be steered to a strongly interacting regime mimicking an effective Fermi-Hubbard model with highly tunable attractive interactions. (b) The atomic level structure.  (c) Coherently mapping the stationary excitations to  propagating photon pulses allows for the efficient probing of the BCS-BEC crossover by measuring the temporal correlations of the photon pulses leaving the fiber.} \label{fig:Fig1}
\end{figure}
%%%%%%%%%%%%%%%%%%%%%%%%%%%%%%%%%%%%%%%%%%%%%%%%%%%%%%%%%%%%%%%%%%%%%%%%%%%%%%%

{\it The system:}
As shown in Fig.~\ref{fig:Fig1}, we consider a waveguide filled with two species, $a$ and $b$, of cold atoms;
\textit{e.g.}, a hollow-core photonic crystal fiber doped with Rubidium isotopes \cite{hollow-core}.
Our scheme involves four hyperfine levels for each atomic specie, two pulsed quantum fields, $\hat{E}_{(1,2)}\left( z,t\right)$, and two classical laser fields, $\Omega _{(a,b)}\left( t\right) $.

Initially, the medium is illuminated by the leftward propagating quantum pulses $\hat{E}_{(1,2),+}\left( z,t\right)$ and classical fields $\Omega _{(a,b),+}\left( t\right) $; henceforth, the plus (minus) subscript will be used to represent leftward (rightward) propagation.
After the quantum pulses $\hat{E}_{(1,2),+}\left( z,t\right)$ completely enter the medium, switching off the  classical fields $\Omega _{(a,b),+}\left( t\right)$ adiabatically converts the quantum pulses into atomic excitations in the usual slow light manner \cite{EIT_review,DPS}.
Then, simultaneously switching on the four classical fields $\Omega _{(a,b),\pm }\left( t\right)$ creates a Bragg grating that traps the quantum fields \cite{EIT_review}. 
In the following, we show how the trapped fields experience strong Kerr nonlinearities induced via their interaction with the atoms, steering the system to a strongly interacting regime described by a two component Lieb-Liniger (LL) model.

In the interaction picture, the Hamiltonian of our system shown in Fig.~1 is
\begin{eqnarray}
H=H^{a}+H^{b},
\end{eqnarray}
with
\begin{eqnarray}
H^{x} &=& -\hbar n^{x} ~\int dz~ \Big\{ \Delta_{2}^{x}\sigma_{22}^{x}+\Delta _{3}^{x}\sigma _{33}^{x}+\Delta _{4}^{x}\sigma _{44}^{x} \nonumber \\
&& + \sqrt{2\pi} \sum_{j=1}^{2} g_{j}^{x}\left( \sigma _{21}^{x}+\sigma _{43}^{x}\right) e^{i \delta^{x}_{j} t} \nonumber \\
&& \times \left( \hat{E}_{j,+}e^{ik_{\mathrm{qu}}^{(j)}z}+\hat{E}_{j,-}e^{-ik_{\mathrm{%
qu}}^{(j)}z}\right)  \nonumber \\
&& + \left[ \left( \Omega _{x,+}e^{ik_{\mathrm{cl}}^{(x)}z}+\Omega _{x,-}e^{-ik_{%
\mathrm{cl}}^{(x)}z}\right) \sigma _{23}^{x} + \mathrm{h.c.} \right] \Big\} \nonumber \\
\end{eqnarray}
where $x=a,b$ denote different atomic species. The continuous collective atomic spin operations, $\sigma _{pq}^{x}\equiv \sigma _{pq}^{x}\left(z,t\right) $, give the average of $\left\vert p\right\rangle ^{x}\left\langle q\right\vert $\
over the $x$-type atoms in a small but macroscopic region around spatial coordinate $z$. The wavevectors are given by $k_{\mathrm{qu}}^{(1,2)}$, $k_{\mathrm{cl}}^{(a,b)}$, and their central frequencies by, $\omega _{\mathrm{qu}}^{(1,2)}$ , $\omega _{\mathrm{cl}}^{(a,b)}$, for the slowly varying quantum fields and classical fields, in that order. For notational simplicity, we omit the space and time dependence of the operators.
We denote the atomic densities $n^{x}$ and $g_{j}^{x}$ denotes the single-photon-single-atom coupling constant between an $x$-type atom and the $j$th quantum field.
Here, we have assumed that the quantum fields $\hat{E}_{(1,2)}$ drive the transitions $\left\vert 2\right\rangle ^{x}\left\langle 1\right\vert $\ and $\left\vert 4\right\rangle ^{x}\left\langle 3\right\vert $ with the same strength $g_{(1,2)}^{x}$.
The one-photon detunings are denoted as $\Delta _{2}^{x}$ and $\Delta _{4}^{x}$; the two-photon detunings as $\Delta _{3}^{x}$; the quantum pulse detunings are written as $\delta^{a}_{j}$ with $\delta^{a}_{2} = \omega_{\mathrm{qu}}^{(1)} - \omega_{\mathrm{qu}}^{(2)}$, $\delta^{b}_{1} = -\delta^{a}_{2}$, and $\delta^{a}_{1} = \delta^{b}_{2}=0$.

Similarly  to earlier slow-light schemes \cite{DPS}, we define polariton operators $\Psi _{j,\pm }=g_{j}^{x_{j}}\sqrt{2\pi n^{x_{j}}}\hat{E}_{j,\pm }/\Omega _{x_{j}}$, where  $x_{1,2} = a,b$, that describe the long lived stationary light-matter excitations of the system. Inserting these into the relevant Maxwell-Bloch equations describing the propagation of the quantum fields in the medium, we find that the stationary polariton operators, $\Psi _{1,2}= ( \Psi _{(1,2),+}+\Psi _{(1,2),-} ) /2$, obey two coupled nonlinear Schr\"odinger equations originating from a two-component Lieb-Liniger Hamiltonian:
\begin{eqnarray}
H &=&\int dz\sum_{j=1}^{2}\Big\{\Psi _{j}^{\dagger }\Big[-\frac{1}{2m_{j}}\nabla
^{2}+V_{0}^{\left( j\right) }+  \nonumber \\
&&V_{1}^{\left( j\right) }\cos ^{2}\left( \pi n_s^{\left(
j\right) }z\right) \Big]\Psi _{j}+\chi _{j}\Psi _{j}^{\dagger }\Psi
_{j}^{\dagger }\Psi _{j}\Psi _{j}\Big\}  \nonumber \\
&&+\chi _{12}\int dz\Psi _{1}^{\dagger }\Psi _{1}\Psi _{2}^{\dagger }\Psi
_{2}.
\end{eqnarray}
Here, the effective masses of polaritons are $m_{j} = - \Delta\omega^{(j)} / (2 \nu^{(j)} v_{g}^{(j)}) - \Gamma^{x_{j}}_{\mathrm{1D}} n^{(x_j)} / (4 \Delta_{2}^{x_{j}} v_{g}^{(j)})$ with $\Delta \omega^{ \left(1,2\right) }=\omega _{\mathrm{qu}}^{(1,2)}-\omega _{\mathrm{cl}}^{(a,b)}$ and $ v_{g}^{(1,2)} = \nu^{(1,2)}\Omega_{(a,b)}^2/\pi(g_{(1,2)}^{(a,b)})^2n^{(a,b)}$ are the group velocities of the quantum fields in the medium with $\nu^{(1,2)}$ the corresponding velocities in an empty waveguide. The spontaneous emission rate of a single atom into the waveguide modes $\Gamma^{x_{j}}_{1D} = 4 \pi (g^{x_{j}}_{j})^2 / \nu^{j}$.
The intra-species repulsions are $\chi _{j} = ( \Lambda^{x_{j}} )^2 \Xi ^{x_{j}} \Gamma^{x_{j}}_{1D} v_{g}^{(j)} / (2 \Delta^{x_{j}}_{4})$ and the interspecies repulsions is $\chi
_{12}= \frac{n^{a} \Gamma _{\mathrm{1D}}^{a} v_{g}^{(1)} ( \Lambda^{a} \Omega _{b} g_{2}^{a} )^{2}}{ 2 n^{b} ( \Delta_{4}^{a} -\delta_2^a) ( \Omega _{a}
g_{1}^{a} )^{2} } + \frac{n^{b} \Gamma _{\mathrm{1D}}^{b} v_{g}^{(2)} ( \Lambda^{b} \Omega _{a} g_{1}^{a} )^{2}}{  2 n^{a} ( \Delta_{4}^{b}  - \delta_1^b ) ( \Omega _{b}
g_{2}^{b} )^{2} } $, where $\Lambda^{x_{j}} = \Omega_{x_{j}}^{2}/(\Omega _{x_{j}}^{2}-\Delta _{3}^{x_{j}}\Delta _{2}^{x_{j}}/2)$ and $\Xi^{x_{j}}=(\Delta _{4}^{x_{j}}-\Delta _{3}^{x_{j}}/2)/(\Delta _{4}^{x_{j}}-\Delta_{3}^{x_{j}})$. The Hamiltonian above contains a periodic potential term induced by a slight modulation in the atomic density distributions for the two atomic species \cite{hollow_recent}. This is possible, for example, by applying two external fields. Thus the atoms, initially prepared in their ground states $\left\vert 1\right\rangle^{(a,b)} $, have densities given by $n^{x_{j}}=n_{0}^{x_{j}}+n_{1}^{x_{j}} \cos ^{2}(\pi n_s^{(x_j)}z)$, where $n_s^{(x_j)}$ is the linear density of sites for atomic specie $x_j$, and $n_{0}^{x_{j}}\gg n_{1}^{x_{j}}$; \textit{i.e.}, the modulation is only a perturbation.
Such periodic distributions for the atomic species give the trapping polaritonic potential $V_{0}^{(j)}= \frac{\Delta \omega^{(j)} v_{g}^{(j)}}{\nu^{(j)}}-\frac{ \Lambda^{x_{j}} \Gamma _{\mathrm{1D}}^{x_{j}} \Delta_{3}^{x_{j}} v_{g}^{(j)} n_{0}^{x_{j}}}{4 \Omega_{x_{j}}^{2}}$ and the periodic polaritonic lattice with depth $V_{1}^{(j) }=-\frac{\Lambda^{x_{j}} \Gamma_{\mathrm{1D}}^{x_{j}} \Delta _{3}^{x_{j}} v_{g}^{(j)} n_{1}^{x_{j}}}{4\Omega _{x_{j}}^{2}}$.

{\it A two component Bose-Hubbard model of polaritons:}
For a sufficiently strong periodic potential and weak interaction between the polaritons, the LL Hamiltonian above can be mapped to a two species Bose-Hubbard model. For simplicity, we assume $ \Delta \omega ^{(j)}=0$ and that the counter propagating classical fields are identical, and the two atomic species have identical mass, distribution, and interaction with the corresponding quantum fields, \textit{i.e.},  $\nu^{(j)}=\nu$, $v_{g}^{(j)}=v_{g}$, $\Omega _{x_{j}} = \Omega $, $n_s^{(j)}=n_s$,
$n^{x_{j}} = n$, $n_{1}^{x_{j}} = n_{1}$, $\Gamma _{\mathrm{1D}}^{x_{j}} = \Gamma _{\mathrm{1D}}$, $g_{j}^{x_{j}}=g$ and $\Delta_{k}^{x_{j}} = \Delta _{k}$\ for $k = 2,3,4$. The conditions on the optical parameters regime for the mapping from LL to BH to be valid  translate to the single photon detuning $\Delta_{4}/\Gamma \ge 20$ and control laser Rabi frequencies of  $\Omega/\Gamma \le 3$.  $\Gamma$ is the atomic decay rate which is assumed at $20\mathrm{MHz}$ for the typical Rb transition in question. The above conditions are calculated following the methods in \cite{hollow_recent}, for atomic densities equal to $n=10^{6}\mathrm{m}%
^{-1}$ with 10$\%$ modulation $n_{1}=0.1n_{a}$; photonic densities of $n_s=10^{2}\mathrm{m}^{-1}$, single atom cooperativity $%
\eta=\Gamma _{\mathrm{1D}}/\Gamma=0.2 $ and two photon detunings $\Delta _{2}=-5\Gamma $, $\Delta_{3} =-0.01\Gamma $ \footnote {We 
note here that these parameters correspond to optical depths of a thousand. Recent experiments have reported a six-fold increase to previously reported values, with  OD $\sim180$ which we hope will make our proposal implementable in the near future (M. Bajcsy \textit{et al.},  arXiv:1104.5220).}

Assuming that the two-photon detuning obeys $|\Delta _{3}| \ll \left\vert \frac{\Omega ^{2}}{\Delta _{2}} \right\vert,\Delta_{4} $, so that the dimensionless quantities $\Lambda^{j},\Xi^{j}\simeq 1$,  we obtain a two component BH model for polaritons:
\begin{equation}
\label{eq:Eq5}
H=-\sum_{\left\langle i,j\right\rangle, \sigma }t_{\sigma } a_{i\sigma
}^{\dagger} a_{j\sigma }+\sum_{i,\sigma }\frac{U_{\sigma }}{2}n_{i\sigma
}^{2}+V\sum_{i}n_{i\uparrow }n_{i\downarrow },
\end{equation}
where $a_{i\sigma}$ is an annihilation operator of a $\sigma$-type polariton at $i$th site and $\langle i,j \rangle$ stands for nearest neighbors. $t=4V_{1}^{3/4}E_{R}^{1/4}\exp \left( -2\sqrt{V_{1}/E_{R}}\right) /\sqrt{\pi }$ $= t_{\uparrow} = t_{\downarrow}$,  $U=\sqrt{2\pi }\chi _{1}n_s\left(
V_{1}/E_{R}\right) ^{1/4} = U_{\uparrow} = U_{\downarrow}$, $V=\sqrt{2\pi }\chi
_{12}n_s\left( V_{1}/E_{R}\right) ^{1/4}/2$, where $E_{R}=\pi
^{2}n_{s}^{2}/(2m)$ with $m=-\Gamma _{\mathrm{1D}}n/\left( 4\Delta
_{2}v_{g}\right) $ and $n_{s}=$ density of sites; $\chi =\Gamma _{\mathrm{1D}}v_{g}/\left( 2\Delta
_{4}\right) =\chi_1=\chi_2$,\ $\chi _{12}=\Gamma _{\mathrm{1D}}v_{g}\Delta _{4}/\left[
\Delta _{4}^{2}-\left( \delta_2^a \right) ^{2}\right] $, and $V_{1}=-\Gamma _{%
\mathrm{1D}}\Delta _{3}v_{g}n_{1}/\left( 4\Omega ^{2}\right) $.

To discuss the simulation of BCS-BEC crossover, we focus on the case of repulsive intra-species ($U>0$), and attractive inter-species ($V<0$) interactions, which can be achieved by setting $\chi >0$\ and $\chi _{12}<0$. The ratios between inter- and intra-species interaction  $V/U$\ $=\chi _{12}/\chi _{1}=\frac{ \Delta_{4}^{2}}{\Delta _{4}^{2}-(\delta_2^a)^2}$ and the hopping to repulsion ratio $t/U=4 V_{1}^{1/2}E_{R}^{1/2}\exp \left( -2\sqrt{V_{1}/E_{R}}\right) /\left(\sqrt{2}\pi \chi _{1}n_s\right)$ determine the physics of the Hamiltonian (\ref{eq:Eq5}) completely. Assuming the effective mass to be positive, the constraints on the one-photon detunings to probe the BCS regime are $\Delta _{2}<0$ and $0<\Delta_{4}<\delta_2^a$. We note here that different regimes are also possible that can lead to different effects such as spin-charge separation or Kondo physics.

\begin{figure}
\includegraphics[width = \columnwidth]{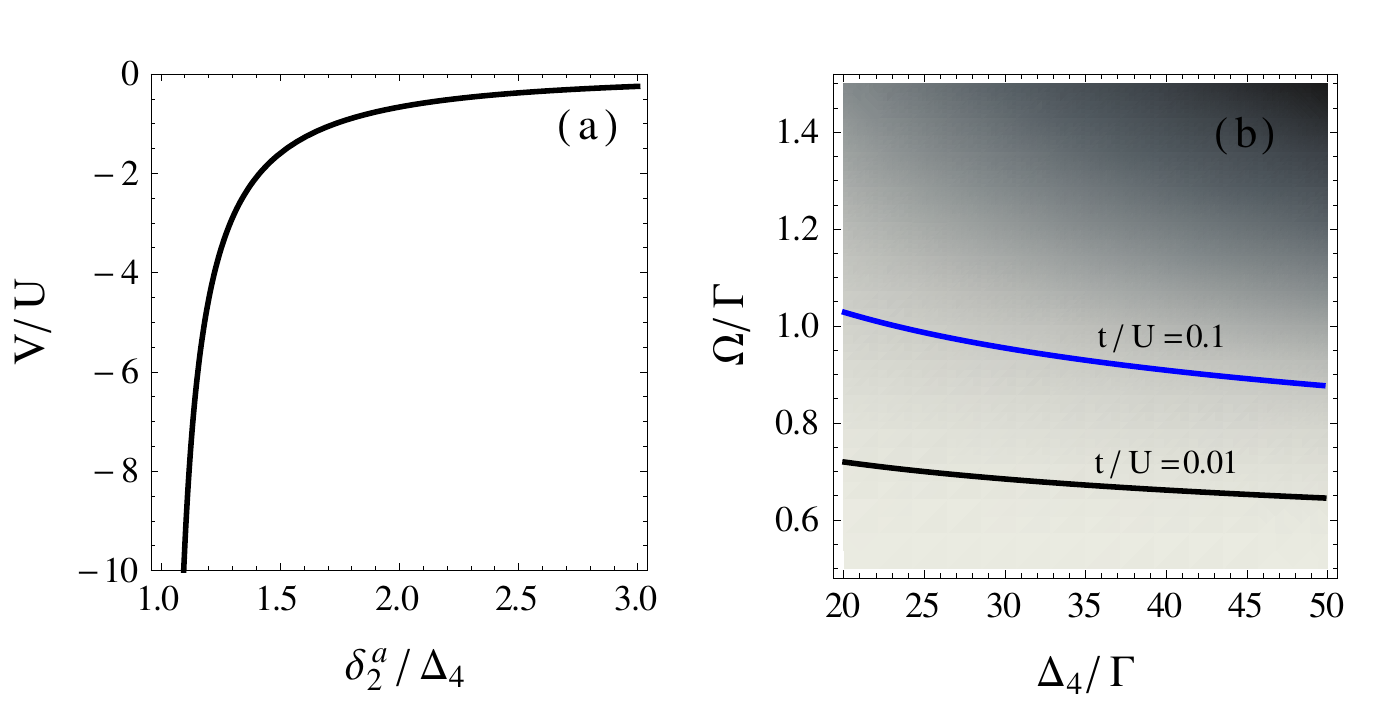} %
\caption{ (a) Inter-species interaction ratio $V/U$ and (b) hopping parameter ratio $t/U$, characterizing the realizable two component BH model of photons as a function of the quantum pulse $\delta^{a}_{2}$ and single photon $\Delta_{4}$ detunings.  $\Gamma _{\mathrm{1D}}=0.2\Gamma $, $n/n_s=10^{4}$, $n/n_{1}=10$, $\Delta
_{2}=-5\Gamma $, and $\Delta _{3}=-0.01\Gamma $.  }\label{fig:Fig2}
\end{figure}

{\it From bosonic polaritons to paired fermions: }
Figure \ref{fig:Fig2} shows the accessible regimes of the inter- and intra-species interactions such that the BH model of polaritons can be mapped into an effective FH model. It is shown that by simply tuning quantum optical parameters such as the single photon detunings $\Delta_{4}$, the detuning between the quantum fields $\delta_2^a$, and strength of the classical trapping lasers $\Omega$, the required regime, $t/U\ll1$, can be reached, while leaving a range of values of $V/U$ accessible. 

%%%%%%%%%%%%%%%%%%%%%%%%%%%%%%%%%%%%%%%%%%%%%%%%%%%%%%%%%%%%%%%%%%%%%%%%%%%%%%%
%%%%%%%%%%%%%%%%%%%%%%%%%%%%%%%%%%%%%%%%%%%%%%%%%%%%%%%%%%%%%%%%%%%%%%%%%%%%%%%

Under these restrictions, the polaritonic two-species Bose-Hubbard model shows the fermion-like BEC-BCS crossover as well as the highly bosonic BB behaviour \cite{Paredes}. Preparing these BCS- and BEC-like states is possible here by simple tuning of the optical parameters, forcing the trapped polaritons to form large Cooper-pair-like objects or localized bosonic molecules, on demand.
Switching off the trapping lasers and coherently mapping the stationary polaritonic correlations to propagating photon pulses allows one to probe the different states by observing the temporal (and hence spatial) second-order correlation functions of the  leaving the waveguide.
We note here the \textit{in situ} character of the measuring process in contrast to the cold atom cases where, usually, time of flight imaging is needed \cite{Bloch_review}.

To discuss the crossover, we focus on the second order cross-species correlations 
$g^{(2)}_{\uparrow \downarrow}(l) = \sum_{i} \langle n_{i \uparrow} n_{i+l \downarrow} \rangle$
and  the  correlations between cross-species population differences
$g_{-}^{(2)}(l) = \sum_{i} \langle (n_{i \uparrow} -  n_{i \downarrow}) (n_{i + l \uparrow} -  n_{i + l \downarrow})\rangle$ as functions of the site number difference $l$. 
 The dependence on $t/U$ and $V/U$ is implied. These types of correlation functions can be measured by collecting the component-resolved photon-counting records and analyzing the collected data. For example, one could use a beam splitter and energy-resolving photon detectors to collect the required data. The strength of our proposal lies in the possibility of directly observing the spatial correlations that are the defining characteristics of the different phases.

%%%%%%%%%%%%%%%%%%%%%%%%%%%%%%%%%%%%%%%%%%%%%%%%%%%%%%%%%%%%%%%%%%%%%%%%%%%%%%%
\begin{figure}[h]
%\psfrag{y}[][]{\footnotesize $g^{(2)}_{\uparrow \downarrow}(l), g_{-}^{(2)}(l)$}
\includegraphics[width= \columnwidth]{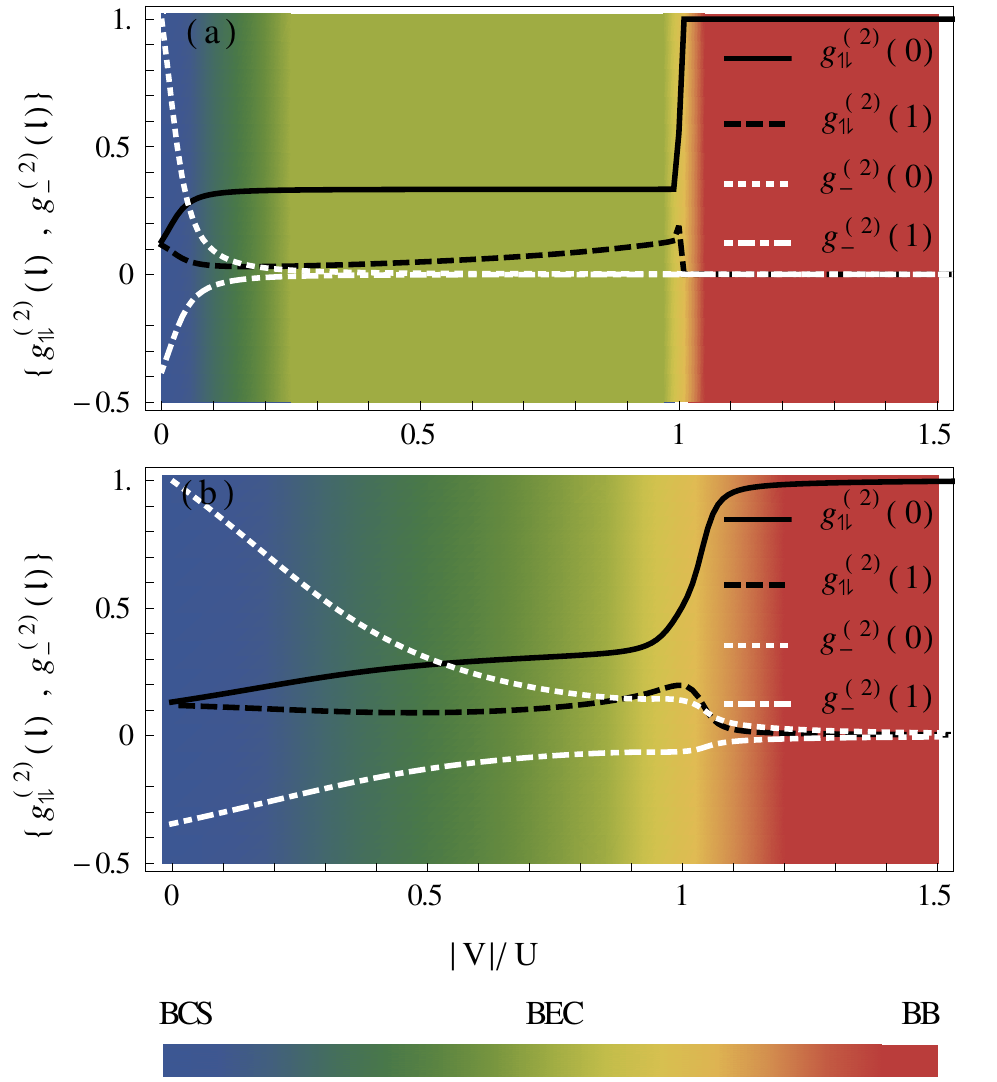}
\caption{Cross-species correlations $g^{(2)}_{\uparrow \downarrow}(l)$ and population difference correlations $g_{-}^{(2)}(l) $ for (a) $t=0.01$ and (b) $t=0.1$ as a function of the inter-species interaction calculated at the same site $l=0$ and for neighbouring sites $l=1$. The colored gradient background, proportionate to the on-site cross-species correlation, portrays the BB-BEC-BCS crossover. By mapping the polaritons to photons and releasing the pulses, the different phases of the system can be probed using photon intensity correlation measurements.}\label{fig:Fig3}
\end{figure}
%%%%%%%%%%%%%%%%%%%%%%%%%%%%%%%%%%%%%%%%%%%%%%%%%%%%%%%%%%%%%%%%%%%%%%%%%%%%%%%

 To calculate the correlation functions, the ground state of Eq.~(4) is computed numerically for 6 polaritons in 8 sites. This corresponds to assuming 3 photons in each quantum field entering the waveguide and a polaritonic potential modulation with the wave vector $k_{m}=2\pi*8/L$, with L the length of the fiber which is taken to be a few centimeters. The BB-BEC-BCS crossover can easily be seen from the on-site correlation $g_{\uparrow\downarrow}^{(2)}(0)$ as shown on Fig.~3. Abrupt changes in $g_{\uparrow\downarrow}^{(2)}(0)$ (normalized to the value at $|V|/U=1.5$) at $|V| = U$ indicate a transition from the big boson (BB) state, where all the polaritons pair up at a single site, to the localized pairing (BEC) state, where different pairs prefer to space out. The curves also indicate a crossover from the locally paired (BEC) state (when $|V| \gg t$) to the long-range paired (BCS) state ($|V| \ll t$).  The coloured background portrays the different phases and how they cross over. The black dashed curves show the correlations at $l = 1$, indicating that the photonic BEC pairs space out in the BEC region as $t/|V|$ is increased until the system crosses over to the BCS regime. The white dotted and dash-dotted curves illustrate $g^{2}_{(-)}(0,1)$ normalized to the value at $|V|/U \rightarrow 0$. They show high sensitivity to the BCS-BEC crossover, but are not suitable for observing the BB-BEC crossover.  In the \textit{in situ} photon correlation measurements, the BB-BEC-BCS crossover will appear as a transition from a highly bunched behaviour in the BB regime, to strongly anti-bunched behaviour in the BSC regime in the on-site cross-species intensity correlations $g_{\uparrow\downarrow}^{(2)}(0)$.

\begin{figure}
%\psfrag{y2}[][]{\footnotesize $g^{(2)}_{\uparrow \downarrow}(l,t/U, V/U)$}
%\psfrag{l}[][]{\footnotesize $l$}
\includegraphics[width= \columnwidth]{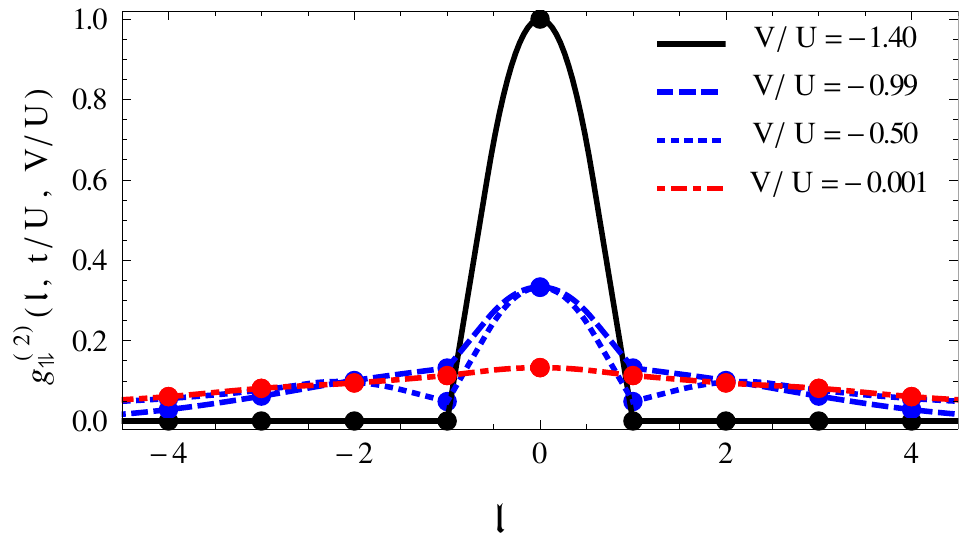}
\caption{Cross-species second order correlation function for $t/U=0.01$ and different values of $|V|/U = 1.2, 0.99,0.5,0.01$ as a function of the distance in units of the effective photonic lattice spacing $l$ .}\label{fig:Fig4}
\end{figure}

While the correlation functions at $l=0,1$ give a good signpost for the three phases, correlations at longer distances are required, especially in the BCS regime, to completely describe the physics. Figure \ref{fig:Fig4} shows $g_{\uparrow\downarrow}^{(2)}(l)$ for different values of $|V|/U$ with $t/U=0.01$ fixed. The expected short-range to long-range crossovers are clearly visible, indicating the measurable signs of the elusive BCS-BEC crossover in our scheme.
%%%%%%%%%%%%%%%%%%%%%%%%%%%%%%%%%%%%%%%%%%%%%%%%%%%%%%%%%%%%%%%%%%%%%%%%%%%%%%%
%%%%%%%%%%%%%%%%%%%%%%%%%%%%%%%%%%%%%%%%%%%%%%%%%%%%%%%%%%%%%%%%%%%%%%%%%%%%%%%

\textit{Conclusion:} We have shown that slow-light-EIT-based techniques can be used to generate a two-species 1D Bose-Hubbard model of photons inside a hollow-core waveguide filled with cold atoms. The tunability of the optical parameters allows for the strongly repulsive regime to be reached, where the predictions of the 1D Fermi-Hubbard model can be studied experimentally. In particular, the resulting strongly correlated polaritons (or photons)  in the waveguide can mimic the BB-BEC-BCS crossover by appropriate tuning of the optical parameters. The possibility of \textit{in situ} measurements of the spatial correlation functions makes our scheme an excellent candidate for an efficient observation of the crossovers using light.

\bibliographystyle{unsrt}

\end{document}